\documentclass[twocolumn,usenatbib]{mn2e}

\usepackage{natbib}

\usepackage{amsmath}
\usepackage{amssymb}
\usepackage{float}
\usepackage{color}
\usepackage{colortbl}
\usepackage{array}
\usepackage{url}

\newcommand{\rx}{{\rm x}}
\newcommand{\ry}{{\rm y}}

\newcommand{\rn}{{\rm n}}
\newcommand{\rp}{{\rm p}}
\newcommand{\re}{{\rm e}}
\newcommand{\rc}{{\rm c}}

\newcommand{\cE}{{\cal E}}

\newcommand{\rL}{{\rm L}}

\newcommand{\cN}{{\cal N}}
\newcommand{\cR}{{\cal R}}

\newcommand{\en}{\varepsilon_\rn}
\newcommand{\ep}{\varepsilon_\rp}
\newcommand{\rd}{{\rm d}}

\newcommand{\cm}{{\rm cm}}
\newcommand{\g}{{\rm g}}
\newcommand{\rs}{{\rm s}}
\newcommand{\K}{{\rm K}}
\newcommand{\yr}{{\rm yr}}
\newcommand{\ms}{{\rm ms}}
\newcommand{\G}{{\rm G}}

\begin{document}

\title[Magnetic Field Evolution in Superconducting Neutron Stars]{Magnetic Field Evolution in \\ Superconducting Neutron Stars}

\author[V. Graber et al.]{Vanessa Graber,$^1$\thanks{E-mail: vanessa.graber@soton.ac.uk}
Nils Andersson,$^1$ Kostas Glampedakis$^{2,3}$ and Samuel K. Lander$^1$ \vspace{0.2cm}\\ 
$^1$Mathematical Sciences and STAG Research Centre, University of Southampton, Southampton SO17 1BJ, United Kingdom \\
$^2$Departamento de Fisica, Universidad de Murcia, 30100 Murcia, Spain \\
$^3$Theoretical  Astrophysics,  University  of  T\"ubingen, Auf  der  Morgenstelle  10, 72076 T\"ubingen, Germany}

%%%%%%%%%%%%%%%%%%%%%%%%%%%%%%%%%%%%%%%%%%%%%%%%%%%%%%%%%%%%%%%%%%%%%%%%%%%%%%%%%%

\maketitle

\begin{abstract}

The presence of superconducting and superfluid components in the core of mature neutron stars calls for the rethinking of 
a number of key magnetohydrodynamical notions like resistivity, the induction equation, magnetic energy and flux-freezing.
Using a multi-fluid magnetohydrodynamics formalism, we investigate how the magnetic field evolution is modified when neutron 
star matter is composed of superfluid neutrons, type-II superconducting protons and relativistic electrons. As an application 
of this framework, we derive an induction equation where the resistive coupling originates from the mutual friction between 
the electrons and the vortex/fluxtube arrays of the neutron and proton condensates. The resulting induction equation allows 
the identification of two timescales that are significantly different from those of standard magnetohydrodynamics. The 
astrophysical implications of these results are briefly discussed.  

\end{abstract}

% Our main task is to 
% % determine how effects acting on the fluxtubes on mesoscopic scales influence the macroscopic field of the star.


%%%%%%%%%%%%%%%%%%%%%%%%%%%%%%%%%%%%%%%%%%%%%%%%%%%%%%%%%%%%%%%%%%%%%%%%%%%%%%%%%%
%%%%%%%%%%%%%%%%%%%%%%%%%%%%%%%%%%%%%%%%%%%%%%%%%%%%%%%%%%%%%%%%%%%%%%%%%%%%%%%%%%

\section{Introduction}
\label{sec-Intro}

Existing at the extremes of physics, neutron stars serve as excellent cosmic laboratories. Some of the most striking features are 
related to their magnetic properties. Measured magnetic field strengths, generally inferred from the star's dipole spin-down, by 
far exceed the strengths of terrestrial magnets. Observations also suggest a link between the various classes of neutron stars. 
Old \textit{millisecond pulsars} thought to be formed in low-mass X-ray binaries have fields between $10^{8}-10^{10} \, \G$, while 
the fields of `classical' \textit{rotation-powered pulsars} range between $10^{10}-10^{13} \, \G$. A third class of slow-rotating, 
highly magnetised neutron stars, so-called \textit{magnetars}, reaches field strengths up to $10^{15} \, \G$. This class is believed 
to include both, soft gamma repeaters and anomalous X-ray pulsars. Understanding the long-term evolution of the stars' magnetic 
fields might be key to establishing connections between the different classes and forming a unified picture of the neutron star 
`zoo' \citep{Kaspi2010, Vigano2013, Harding2013}. 

Unsurprisingly, such enormous strengths suggest that magnetic fields are crucial for the neutron stars' dynamics. As first pointed 
out by Thompson \& Duncan (\citeyear{Thompson1995, Duncan1996}), magnetic field decay on a timescale of $\sim 10^4 \, \yr$ could 
power the high activity of magnetars. The rotational energy is not sufficient to explain the observed emission of these objects. 
There are also observations indicating that the magnetic dipole fields of standard pulsars evolve on a timescale of the order $10^7 
\, \yr$ \citep{Lyne1985, Narayan1990}. The actual mechanisms, causing the magnetic field to change on these rather short timescales, 
are only poorly understood and there is no definitive answer to the question of which part of the neutron star dominates the magnetic
field evolution. Most theoretical studies and numerical simulations focus on the crust as the source of the field decay and 
neglect the core contribution \citep{Pons2007, Vigano2013, Gourgouliatos2014}. However, one could argue that the core, which 
carries the majority of the star's inertia and magnetic energy, should also play a role in the magnetic field evolution. 

The problem of magnetic field evolution in isolated neutron stars has been discussed by a number of authors. \citet{Goldreich1992}
determined several mechanisms that are present in an ionised plasma consisting of neutrons, protons and electrons. \textit{Ohmic 
diffusion} due to the interaction of relativistic electrons and lattice nuclei causes magnetic field dissipation in the crust. This 
mechanism is most effective on small scales and, thus, not expected to affect the large scale evolution of the crustal field. 
However, Ohmic decay could be enhanced by \textit{Hall drift}, which is in itself conservative but may redistribute magnetic energy
from large to gradually smaller lengthscales. The combined effect, sometimes referred to as the Hall cascade, could cause field 
evolution on a timescale of order $10^7$ years \citep{Goldreich1992}. However, recent neutron star crust simulations show no strong 
cascading behaviour but suggest the existence of a quasi-equilibrium established on timescales shorter than the Ohmic timescale
\citep{Pons2010, Gourgouliatos2014, Wood2015}. In the core, standard Ohmic decay is negligible because the interior is expected to 
form a type-II superconductor (\citeauthor{Baym1969} \citeyear{Baym1969}, see also below). Electron-proton scattering, already acting 
on very long timescales in the star's interior \citep{Baym1969b}, is then 
restricted to the normal conducting cores of fluxtubes. These only contribute a tiny fraction to the star's total cross section. 
Hence, the coupling timescale is increased further, making this dissipation mechanism irrelevant. Additionally, \textit{ambipolar
diffusion}, describing the motion of charged particles and magnetic field lines relative to the neutrons, could cause magnetic 
field decay and drive the flux from the core to the crust \citep{Goldreich1992}. This mechanism was originally considered by 
\citet{Thompson1995} to explain the magnetar activity. However, more recent results seem to indicate that the timescale for 
ambipolar diffusion considerably increases when the superfluid nature of the neutrons or proton superconductivity is taken into 
account \citep{Glampedakis2011c}.

Lacking a clear answer, the question of magnetic field evolution in a neutron star is revisited in this paper. We focus on the outer
core and use the formalism developed by \citet{Glampedakis2011a} (see also \citeauthor{Mendell1991} \citeyear{Mendell1991}, 
\citeauthor{Mendell1991a} \citeyear{Mendell1991a}), which presents a general set of macroscopic hydrodynamic equations for a 
multi-fluid mixture. We address, in particular, the effect the superconducting component has on the evolution of the magnetic field. 

The presence of superfluid and superconducting components in neutron stars is firmly supported by observations and microphysical
calculations. Traditionally, glitches and post-glitch relaxation timescales on the order of months to years are seen as 
observational evidence of superfluidity. \citet{Anderson1975} first proposed that the neutron star's dynamical evolution during 
and after a glitch could be explained by the weak viscous properties of a superfluid component that is coupled to the crust.
Moreover, recent spectral analyses of the neutron star in the supernova remnant Cassiopeia A indicate that the surface temperature 
of this young object decreases faster than one would expect from standard cooling models \citep{Page2011, Shternin2011}. The rapid
cooling could be explained by enhanced neutrino emission, resulting from the onset of neutron superfluidity and proton superconductivity 
in the core. In addition to observations, theoretical calculations provide strong reasons for macroscopic quantum condensates in 
neutron star cores. The idea of superfluid interiors was first put forward by \citet{Migdal1959}, several years before the first 
detection of a pulsar \citep{Hewish1968}. A few hundred years after birth, neutron stars are in thermal equilibrium and have 
temperatures of $10^6 - 10^8 \, \K$ \citep{Tsuruta1998, Page2006, Ho2012b}. While this is certainly hot in terms of terrestrial physics, 
the temperatures lie well below the Fermi temperature of nuclear matter, which is of the order $10^{12} \, \K$ \citep{Sauls1989}. 
Applying the microscopic theory of laboratory superconductors \citep{Bardeen1957} to the neutron star interior suggests that the neutrons 
and protons form Cooper pairs and, thus, condense into a superfluid and a superconductor, respectively.

The properties of the proton superconductor were further discussed in the seminal paper by \citet{Baym1969}. As the conductivity
of normal conducting matter is very large \citep{Baym1969b}, the authors argued that flux could not be expelled from the star's 
interior. The phase transition into a superconducting state has to occur at constant magnetic flux. Two characteristic 
lengthscales determine the state of the superconducting protons. The \textit{London penetration depth} describes the magnetic
field's exponential fall-off from the surface of a superconductor or the fluxtube cores due to the Meissner effect. The 
\textit{coherence length}, which characterises the typical dimensions of a Cooper pair, is equivalent to the core radius of a
fluxtube. The ratio of these two quantities, the Ginzburg-Landau parameter, $\kappa$, dictates the type of superconductivity. For 
$\kappa > 1/ \sqrt{2}$, it is energetically favourable for the material to increase the surface area between normal and 
superconducting regions, implying that magnetic flux can penetrate the medium in the form of quantised fluxlines 
\citep{Abrikosov1957}. \citet{Baym1969} estimated $\kappa$ for the neutron star's outer core and found that the protons would 
enter a metastable type-II state by forming a regular array of fluxtubes (see also Figure 1 of \citeauthor{Glampedakis2011a} 
\citeyear{Glampedakis2011a}). 

% - type-II state even fields below $B\approx H_{\rm c1}$ 

The presence of fluxtubes significantly influences the magnetic properties of the star because the flux is no longer locked to the
charged plasma but is mainly confined inside the fluxtubes. As standard coupling mechanisms, like Ohmic dissipation, are suppressed 
as a result of pairing, interactions of fluxlines with their surroundings determine the magnetic field evolution on macroscopic 
lengthscales. The most prominent of these effective coupling processes, known as mutual friction, is the scattering of electrons 
off the fluxtube magnetic field \citep{Alpar1984, Mendell1991a, Andersson2006b}. In the following, we derive an equation for the
magnetic field evolution of a superfluid/superconducting mixture using a smooth-averaged formalism. We translate the mesoscopic
phenomena, influencing individual fluxtubes, to the large scale picture by applying the framework known from standard resistive 
magnetohydrodynamics (MHD).

The structure of this paper is as follows. In Section \ref{sec-Background}, we introduce the hydrodynamical equations for a 
multi-fluid system and the corresponding Maxwell equations. In Section \ref{sec-StandardMHD}, the standard resistive MHD formalism
is reviewed. We then discuss the case of superconducting matter in Section \ref{sec-SuperconMHD}, focussing on the standard 
resistive coupling introduced by \citet{Alpar1984}. After giving the most general form of the superconducting induction equation, 
we use a few assumptions to simplify and interpret our results. We finally conclude with a discussion in Section 
\ref{sec-Discussion}. 

Note that we will work in an inertial frame and carry out the analysis in a coordinate basis. Vectors are, thus, denoted by their 
individual components. We also use Gaussian units in the remainder of this paper.

% - Gabler QPO simulations, need a core magnetic field to get to the frequencies, crustal field is not enough
% - most simulations consider just the crust and take for simplicity the core as a flux free, hence type-I regime
% - mainly for the lack of having an alternative description, matching both would allow to model the coupled system
% and maybe get to something like the dynamical evolution of glitches

%%%%%%%%%%%%%%%%%%%%%%%%%%%%%%%%%%%%%%%%%%%%%%%%%%%%%%%%%%%%%%%%%%%%%%%%%%%%%%%%%%
%%%%%%%%%%%%%%%%%%%%%%%%%%%%%%%%%%%%%%%%%%%%%%%%%%%%%%%%%%%%%%%%%%%%%%%%%%%%%%%%%%

\section{Theoretical Background}
\label{sec-Background}

\subsection{Multi-fluid hydrodynamics}
\label{subsec-MultifluidHydro}

The model presented in this section largely builds upon the recent work by \citet{Glampedakis2011a}. Making use of the 
Lagrangian formalism developed by Carter, Prix and collaborators \citep{Carter1995a, Prix2004, Andersson2006}, a full 
set of MHD equations for the superfluid/superconducting bulk in the outer neutron star core is derived. The simplest 
representation of this Fermi liquid is a mixture of three components, namely relativistic electrons, superconducting 
protons and superfluid neutrons. In the following, the constituents are denoted by roman indices, $\rx =\{\re, \rp, 
\rn \}$.  Note that in order to keep the discussion clear, we neglect the presence of muons. However, generalising to 
the four-constituent case would be straightforward as electrons and muons are strongly coupled and move as one component 
on macroscopic lengthscales \citep{Mendell1991a}.

The dynamics of the three-fluid system is governed by two Euler equations, one representing the superfluid neutrons 
and the other representing the electron-proton conglomerate. The charged fluids can be characterised as a single component 
if charge neutrality holds over macroscopic distances, i.e. $n_\re=n_\rp$, where $n_\rx$ denotes the particle number density. 
As the electrons are mobile enough to quickly equilibrate any local charge imbalances, this requirement is fulfilled in
the neutron star core \citep{Jackson1999, Glampedakis2011a}. Additionally, the electron mass is significantly smaller
than the proton mass, $m_\re \ll m_\rp$, which allows us to neglect any electron inertial terms. The resulting macroscopic
Euler equations are
\begin{align}
	\left[\partial_t + v^j_\rn \nabla_j \right] & ( v^i_\rn + \varepsilon_\rn w^i_{\rm pn} ) -
		\en w_{\rm pn}^j \nabla^i v_j^\rn \nonumber \\[1.3ex]
		&= - \nabla^i \left( \tilde{\mu}_\rn  + \Phi \right ) + f^i_{\rm mf} + f^i_{\rm mag,n},
			\label{eulern_final} \\[2ex]
	\left[ \partial_t + v^j_\rp \nabla_j \right] & ( v^i_\rp + \varepsilon_\rp w^i_{\rm np} ) 
		+ \ep w_{\rm np}^j \nabla^i v_j^\rp \nonumber \\[1.3ex]
		&= - \nabla^i \left ( \tilde{\mu} + \Phi \right ) - \frac{\rho_\rn }{\rho_\rp} f^i_{\rm mf} + f^i_{\rm mag,p}.
			\label{eulerp_final}
\end{align}
The averaged fluid velocities are denoted by $v^i_\rx$ and the mass densities by $\rho_\rx= m  n_\rx$, where $m \equiv m_\rn=m_\rp$
is the baryon mass. $w_\mathrm{xy}^i \equiv v_\rx^i-v_\ry^i$ is the relative velocity, $\Phi$ the gravitational potential and 
$\varepsilon_\rx$ the entrainment parameters. By definition, the latter satisfy the condition $n_\rp \varepsilon_\rp = n_\rn 
\varepsilon_\rn$. The specific chemical potentials are defined as
\begin{equation}
    \tilde{\mu}_\rn \equiv \frac{\mu_\rn}{m}, \qquad \tilde{\mu} \equiv \frac{\mu_\rp + \mu_\re}{m}.
\end{equation}
The two Euler equations are supplemented by three continuity equations for the number densities,
\begin{equation}
    \partial_t n_\rx + \nabla_i \left( n_\rx v_\rx^i \right) = 0,
		\label{eqn-Continuity}
\end{equation}
which reflect the conservation of mass for each individual species, and the Poisson equation
\begin{equation}
    \nabla^2 \Phi = 4 \pi G \rho.
\end{equation}
$\rho=\sum_\rx \rho_\rx$ is the total mass density of the system and $G$ the gravitational constant. 

The variational approach used to derive the Euler and continuity equations explicitly distinguishes between the fluid momenta 
and velocities. This formalism provides the possibility to include any changes, caused by the superfluid and superconducting 
condensates, into the hydrodynamical model. In contrast to the momentum equations of standard plasma physics \citep{Jackson1999},
Equations \eqref{eulern_final} and \eqref{eulerp_final} incorporate new inertial terms due to entrainment, which arise from 
the strong coupling of Fermi liquids, and terms that go beyond the standard electromagnetic interaction given by the Lorentz 
force. For the fluid mixture in the outer neutron star core, the right-hand sides of the momentum equations contain the total 
magnetic and mutual friction forces per unit volume, $f^i_{\rm mag,x}$ and $f^i_{\rm mf}$, respectively. The former one is 
caused by interactions of the vortex/fluxtube magnetic field with the charged fluid. We point out that the neutron fluid 
experiences this magnetic force because protons are entrained around each neutron vortex and create an effective magnetic field.
In the absence of entrainment, the magnetic force on the neutron component would vanish. Finally, the mutual friction forces arise 
from the dissipative coupling of the vortex and fluxtube array with the fluid components.

We keep in mind that our hydrodynamic model is solely based on averaged quantities and reflects the macroscopic behaviour of the 
fluid components. It is on these large scales that we have a method to deal with the presence of the quantum condensates in a 
consistent way. Taking advantage of the large numbers of vortices/fluxtubes, we \mbox{average} over the respective arrays 
and obtain a smooth-averaged description of the magnetic and mutual friction forces. Using this formalism, it is possible to 
determine how the presence of vortices and fluxtubes influences the macroscopic dynamics of a neutron star, i.e. its rotational 
and magnetic evolution. If individual vortices/fluxtubes do not overlap and are distant enough so that interactions within one 
array can be neglected, the averaging procedure is obtained from the macroscopic quantisation conditions originally developed by 
\citet{Onsager1949} and \citet{Feynman1955} for the dynamics of rotating superfluid helium. Assuming that neutron vortices 
and proton fluxtubes are locally straight and directed along the unit vectors, $\hat{\kappa}^i_\rn$ and $\hat{\kappa}^i_\rp$, 
the arrays can be assigned vortex and fluxtube surface densities, $\cN_\rn$ and $\cN_\rp$, respectively. As the vorticities, 
$\mathcal{W}^i_\rx$, are related to the circulation of the averaged canonical momenta, the macroscopic quantisation conditions are 
given by
\begin{align}
	\mathcal{W}_\rn^i &= \epsilon^{ijk} \nabla_j \left( v_k^\rn + \varepsilon_\rn w_k^{\rm pn} \right) = \cN_\rn \kappa^i_\rn , 
		\label{eqn-Quantisation1}  \\[1.4ex]
	\mathcal{W}_\rp^i & = \epsilon^{ijk} \nabla_j \left( v_k^\rp + \varepsilon_\rp w_k^{\rm np} \right) + a_\rp B^i = \cN_\rp \kappa^i_\rp, 
		\label{eqn-Quantisation2}
\end{align}
where $\kappa^i_\rx= \kappa \hat{\kappa}^i_\rx$ points along the local vortex direction with the quantum of circulation
\begin{equation}
	\kappa = \frac{h}{2 m} \approx 2.0 \times 10^{-3} \, \cm^2 \, \rs^{-1},
\end{equation}
and we define
\begin{equation}
	a_\rp \equiv \frac{e}{m c} \approx 9.6 \times 10^3 \, \G^{-1} \, \rs^{-1}.
\end{equation}
The proton charge, the speed of light and the Planck constant are denoted by $e$, $c$ and $h$, respectively. 

\subsection{Macroscopic magnetic induction}
\label{subsec-Induction}

In the averaged framework, the total magnetic induction, $B^i$, is the sum of three individual components, namely the averaged 
fluxtube and vortex field and the London field,
\begin{equation}
	B^i = B_\rp^i + B_\rn^i + b_{\rm L}^i .
		\label{eqn-AveragedField}
\end{equation}
The former two contributions are obtained by multiplying the surface densities, $\cN_\rx$, with the flux carried by a single 
line of the lattice, $\phi_\rx$. The fluxes can be derived by considering the dynamics of a single vortex/fluxtube on mesoscopic
lengthscales, denoted by bars on the respective quantities. Using the corresponding quantisation condition and the mesoscopic 
Amp\`ere law (e.g. see Appendix A1 of \citet{Glampedakis2011a} for details), it is possible to derive generalised London equations 
for the mesoscopic magnetic fields, $\bar{B}_\rx^i$,
\begin{equation}
	  \lambda^2_* \nabla^2 \bar{B}_\rx^i - \bar{B}_\rx^i = - \phi_\rx \hat{\kappa}_\rx^i \delta(\vec{r}\, ),
		  \label{eqn-LondonModified}
\end{equation}
where $\delta(\vec{r}\, )$ is the two-dimensional delta function located at the centre of each vortex/fluxtube, $\phi_\rx$ is
defined below and the effective London penetration depth is given by
\begin{equation}
	\lambda_* = \left(\frac{1}{4 \pi \rho_\rp a_\rp^2} \, \frac{1 -\en-\ep}{1-\en} \right)^{1/2}.
		\label{eqn-LondonDepth}
\end{equation}
In the absence of entrainment, $\en = \ep=0$, this expression reduces to the standard result of superconductivity 
\citep{Tinkham2004}. Taking advantage of the symmetry and using cylindrical coordinates, the inhomogeneous Helmholtz equation
\eqref{eqn-LondonModified} can be solved using a Green's function approach in two dimensions \citep{Fetter1969}. Integrating 
the resulting magnetic induction over a disc of radius $r \gg \lambda_*$ perpendicular to $\hat{\kappa}_\rx^i$ gives for the magnetic
flux
\begin{equation}
	 \int \bar{B}_\rx^i \, \rd S = \phi_\rx \hat{\kappa}_\rx^i.
\end{equation}
For a proton fluxtube, we obtain the expected unit of flux,
\begin{equation}
	\phi_\rp = \phi_0 = \frac{\kappa}{a_\rp} = \frac{hc}{2e} \approx  2.1 \times 10^{-7} \,\G \, \cm^2,
\end{equation}
whereas the flux of a superfluid vortex is 
\begin{equation}
	\phi_\rn = - \frac{\ep}{1-\en} \, \phi_0.
\end{equation}
We note that the minus sign originates from $\hat{\kappa}_\rn^i$ and $\bar{B}_\rn^i$ pointing into opposite directions. In the 
absence of entrainment, the neutron flux would be zero. The averaged contributions from the two arrays to the macroscopic magnetic
induction, $B^i$, are then obtained by
\begin{equation}
	B_\rx^i = \cN_\rx \phi_\rx \hat{\kappa}_\rx^i.
		\label{eqn-BAveraged}
\end{equation}

The third contribution to the induction is the London field. It is a fundamental property of a superconductor and 
associated with its rotation \citep{Tilley1990}. While superfluids need to form vortices in order to support any circulation,
the superconducting fluxtube array is not related to the macroscopic rotation. However, these dynamics induce an additional 
magnetic field inside the superconductor, whose axis is parallel to the rotation axis. In contrast to $B_\rx^i$, the London 
field is not of microscopic origin but related to the macroscopic electromagnetic current (see Subsection \ref{subsec-Maxwell}). 
Combining the quantisation conditions \eqref{eqn-Quantisation1} and \eqref{eqn-Quantisation2} with Equation \eqref{eqn-BAveraged}, 
the London field can be related to the macroscopic fluid properties. Assuming that the hydrodynamical lengthscales are 
sufficiently small to ensure constant entrainment parameters, we have
\begin{equation}
	b_\rL^i = - \frac{1}{a_\rp} \, \frac{1 - \en - \ep}{1-\varepsilon_\rn} \, 
		\epsilon^{ijk} \nabla_j v^{\rm p}_k.
		\label{eqn-LondonField2}
\end{equation}
The proton entrainment parameter is related to the effective proton mass via 
\begin{equation}
	\ep = 1 - \frac{m_\rp^*}{m} \approx 0.3,
		\label{eqn-envalue}
\end{equation}
where we have used the estimate $m_\rp^* \approx (0.6-0.9)m$ given by \citet{Chamel2006} for the outer neutron star core. For 
small proton fractions, $x_\rp \equiv \rho_\rp/\rho \ll 1$, an approximation which is valid in the interior of a neutron star, 
the neutron entrainment coefficient is negligible because $\varepsilon_\rn \approx x_\rp \varepsilon_\rp \ll 1$. Taking the proton
fluid to be tightly coupled to the neutron star's crust through the magnetic field and, thus, rotating rigidly at the observable
pulsar frequency, we can substitute a canonical rotation period to calculate an estimate for the magnitude of the London field. 
Using the normalisation $P_{10} \equiv P/(10 \, \ms)$, we find 
\begin{equation}
	b_{\rm L} \approx 9.2 \times 10^{-2} P^{-1}_{10} \, \G,
\end{equation}
which is many orders of magnitude smaller than the magnetic field strengths usually invoked for neutron star physics. Hence, it 
is generally justified to neglect the London field in Equation \eqref{eqn-AveragedField}, an approach we will take in Subsection
\ref{subsec-SimplifiedSet}.

We can simplify the magnetic induction, $B^i$, one step further by taking the properties of the vortex and fluxtube arrays into 
account. Although the individual fluxes, $\phi_\rx$, are comparable, the contribution from the superconducting protons dominates:
Consider smooth-averaged fluid velocities of the form $\epsilon^{ijk} \nabla_j v_k^\rx = 2 \Omega_\rx \hat{z}^i$. In this case, the 
vortices/fluxtubes are aligned with the $z$-axis, i.e. $\hat{z}^i = \hat{\kappa}^i_\rx$, and Equation \eqref{eqn-Quantisation1} 
gives
\begin{equation}
	\cN_\rn = \frac{2}{\kappa} \left[ \Omega_\rn + \epsilon_\rn \Omega_{\rm pn} \right].	
\end{equation}
As we expect the lag, $\Omega_{\rm pn}=\Omega_\rp - \Omega_\rn$, to be small on macroscopic scales and the neutrons to be coupled 
to the crust, the vortex surface density is
\begin{equation}
	\cN_\rn \approx \frac{4 \pi}{\kappa P} \approx 6.3 \times 10^5 P^{-1}_{10} \, \cm^{-2},
\end{equation}
The neutron vortex density is, thus, fixed by the rotation of the neutron star.
Using previous estimates for the entrainment parameters, the magnetic field strength of the neutron vortex array is
\begin{equation}
	B_\rn \approx \cN_\rn \ep \phi_0 \approx 4.0 \times 10^{-2} P^{-1}_{10} \, \G.
\end{equation}
This is again many orders of magnitude smaller than typical neutron star field strengths, which implies that the magnetic field of 
the outer neutron star core is mainly confined to the proton fluxtube cores. It is, therefore, important to investigate which 
mechanisms affect the motion of individual fluxtubes in order to link the small scale behaviour to the large scale evolution of 
the star's magnetic field. The fluxtube density can be estimated to
\begin{equation}
	\cN_\rp = \frac{B_\rp}{\phi_0} \approx \frac{B}{\phi_0} 
		\approx 4.8 \times 10^{18} \, B_{12} \, \cm^{-2}
		\label{eqn-FluxtubeDensity},
\end{equation}
with the normalised magnetic field $B_{12} \equiv B/(10^{12} \, \G)$.

\subsection{Macroscopic Maxwell equations}
\label{subsec-Maxwell}

In order to capture the electromagnetic response of the fluid mixture correctly, the Euler equations \eqref{eqn-Quantisation1} and 
\eqref{eqn-Quantisation2} have to be supplemented by Maxwell's equations. Taking these to be valid in our multi-fluid mixture, we
have to redefine, or rather reinterpret, the various fields accordingly in order to make Maxwell's equations suitable for a type-II 
superconductor. 

As mentioned before, the London field, despite being of small magnitude, plays an important role for the electrodynamics. As 
discussed by \citet{Carter2000} and \citet{Glampedakis2011a}, the London field is closely connected to the macroscopic 
electromagnetic current,
\begin{equation}
	 J^i = e n_\re w_{\rm pe}^i,
		\label{eqn-Current}
\end{equation}
which enters the macroscopic Amp\`ere law. In contrast to standard MHD, where the equality $H^i=B^i$ is satisfied, the averaged 
magnetic induction, $B^i$, and the macroscopic magnetic field, $H^i$, are no longer equivalent in a type-II superconducting 
sample. Instead, Amp\`ere's law reads
\begin{equation}
	 \epsilon^{ijk} \nabla_j H_k = \epsilon^{ijk} \nabla_j b^\rL_k = \frac{4 \pi}{c} J^i,
		 \label{eqn-AmpereLaw}
\end{equation}
where the displacement current has been neglected because the fluid motion is slow compared to the speed of light. This deviation
from standard MHD can be also understood in terms of the classification generally applied to terrestrial superconductors (see for
example \citeauthor{Tinkham2004} \citeyear{Tinkham2004}). In laboratory experiments, one distinguishes between macroscopic
electromagnetic currents that generate a macroscopic field, $H^i$, and magnetisation currents \textit{only} affecting the 
mesoscopic induction, which is $\bar{B}^i$ in our notation. A supercurrent, circulating around each vortex/fluxtube and 
generating $\bar{B}^i_\rx$, is attributed to the second class. It does not contribute to the field $H^i=b^i_\rL$, which is created
by the current $J^i$. Hence, the macroscopic magnetic induction, $B^i$, given in Equation \eqref{eqn-AveragedField} differs from 
the magnetic field, $H^i$. For comparison, in vacuum or normal conductors, no magnetisation currents are present and the identification 
$H^i=B^i=\bar{B}^i$ can be made. In the present case, 
\begin{equation}
	 \epsilon^{ijk} \nabla_j H_k = \epsilon^{ijk} \nabla_j B_k = \frac{4 \pi}{c} J^i.
		 \label{eqn-AmpereLawStandard}
\end{equation}
In addition to Amp\`ere's law, we use 
\begin{equation}
	\nabla_i B^i =0
		\label{eqn-MaxwellB},
\end{equation}
which has to hold everywhere in the superconducting fluid, and the macroscopic Faraday law,
\begin{equation}
	\partial_t B^i = - c \, \epsilon^{ijk} \nabla_j E_k.
		\label{eqn-Faraday}
\end{equation}
Instead of defining the macroscopic electric field as the average over the microscopic equivalent, we take advantage of the
remaining fluid degree of freedom, namely the electron Euler equation, to obtain an expression for $E^i$. Neglecting again the
electron inertial terms, we have
\begin{equation}
	E^i = -\frac{1}{c} \epsilon^{ijk} v_j^\re  B_k 
		- \frac{m_\re}{e} \nabla^i \left ( \tilde{\mu}_\re + \Phi \right ) - \frac{F^i_\re}{c a_\rp \rho_\rp},
		\label{eqn-ElectricField}
\end{equation}
where $F^i_\re$ represents the total force exerted on the electrons due to interactions with the surrounding fluid components. 

Combining Equations \eqref{eqn-Faraday} and \eqref{eqn-ElectricField} leads to an evolution equation for the magnetic induction 
that only depends on macroscopic fluid variables. However, the procedure \mbox{relies} on the forces, $F^i_\re$, and we address 
this in the following sections.

%%%%%%%%%%%%%%%%%%%%%%%%%%%%%%%%%%%%%%%%%%%%%%%%%%%%%%%%%%%%%%%%%%%%%%%%%%%%%%%%%%
%%%%%%%%%%%%%%%%%%%%%%%%%%%%%%%%%%%%%%%%%%%%%%%%%%%%%%%%%%%%%%%%%%%%%%%%%%%%%%%%%%

\section{Magnetic field evolution in standard MHD}
\label{sec-StandardMHD}

Before discussing the more complicated problem of magnetic field evolution in a superfluid/superconducting mixture, we briefly 
review the approach taken in normal resistive matter, which allows us to compare our new results with a well studied model. 

\subsection{The MHD induction equation}
\label{subsec-MHDInduction}

In a charged plasma containing electrons and protons, the only relative flow present is the motion between the two components. 
Assuming that a frictional mechanism would damp these dynamics and try to bring the two species into co-motion is straightforward.
Hence, resistive coupling acting on a timescale $\tau_\re$ leads to a dissipative force on the electron fluid,
\begin{equation}
	 F_\re^i = \frac{n_\re m_\re}{\tau_\re} w_{\rm pe}^i = - \frac{m_\re}{e \tau_\re} J^i.
		\label{eqn-ForceMHD}
\end{equation}
Substituting this force into Equation \eqref{eqn-ElectricField} gives a generalised Ohm's law
\begin{equation}
	E^i = -  \frac{1}{c} \, \epsilon^{ijk} \left( v_j^\rp - \frac{J_j}{e n_\re} \right) B_k 
		- \frac{m_\re}{e} \nabla^i \left ( \tilde{\mu}_\re + \Phi \right ) + \frac{J^i}{\sigma_\re}
		\label{eqn-OhmsLaw}
\end{equation}
with the standard electrical conductivity defined by
\begin{equation}
	\sigma_\re \equiv \frac{c \rho_\rp a_\rp e \tau_\re}{m_\re} = \frac{n_\re e^2 \tau_\re}{m_\re}.
\end{equation}
Equation \eqref{eqn-OhmsLaw} combined with Faraday's law \eqref{eqn-Faraday} leads to an evolution equation for the 
magnetic induction,
\begin{align}
	 \partial_t B^i &= \epsilon^{ijk} \nabla_j \epsilon_{klm} \left( v_\rp^l B^m \right) 
		- \epsilon^{ijk} \nabla_j \epsilon_{klm} \left( \frac{c^2}{4 \pi \sigma_\re} \nabla^l B^m \right) \nonumber \\[1.6ex]
		&- \epsilon^{ijk} \nabla_j \epsilon_{klm} \left[ \frac{m c}{4 \pi e \rho_\rp} \, \epsilon^{lsp} \left( 
		\nabla_s B_p \right) B^m \right].
		\label{eqn-NormalMHDInduction}
\end{align}
We have used Amp\`ere's law for normal matter \eqref{eqn-AmpereLawStandard} to eliminate the macroscopic current in 
the last expression. The second and the third term on the right-hand side represent the Ohmic decay and the Hall evolution. 
We can extract the following well-known timescales,
\begin{equation}
	 \tau_{\rm Ohm} = \frac{4 \pi \sigma_\re L^2}{c^2}
\end{equation}
and 
\begin{equation}
	 \tau_{\rm Hall} = \frac{4 \pi e \rho_\rp L^2}{m c B},
\end{equation}
where $L$ is the characteristic lengthscale over which the magnetic field changes. 

\subsection{Flux-freezing in MHD}
\label{subsec-MHDFreezing}

We can estimate the two characteristic timescales for a neutron star core. According to \citet{Baym1969b},
the electrical conductivity associated with the interaction of highly relativistic electrons and normal, degenerate protons is 
given by
\begin{equation}
	\sigma_\re \approx 5.5 \times 10^{28} \, T_8^{-2} \, \rho_{14}^{3/2} \left(\frac{x_\rp}{0.05} \right)^{3/2} \, \rs^{-1},
\end{equation}
where $T_8 \equiv T/(10^8 \, \K)$ is the star's normalised temperature, $\rho_{14} \equiv \rho/( 10^{14} \, \g \, \cm^{-3})$ the 
normalised total density and $x_\rp$ the proton fraction. Approximating the characteristic lengthscale by the radius of the 
neutron star, the Ohmic diffusion timescale is 
\begin{equation}
	 \tau_{\rm Ohm} \approx 2.4 \times 10^{13} \, T_8^{-2} \, L_6^2 \, \rho_{14}^{3/2} \left(\frac{x_\rp}{0.05} \right)^{3/2} \, \yr,
		\label{eqn-Ohmicestimate}
\end{equation}
with the normalised lengthscale $L_6 \equiv L/(10^6 \, \cm)$. For the Hall timescale, we obtain
\begin{equation}
	 \tau_{\rm Hall} \approx 1.9 \times 10^{10} \, B_{12}^{-1} \, L_6^2 \, \rho_{14} \left(\frac{x_\rp}{0.05} \right) \, \yr.
		\label{eqn-Hallestimate}
\end{equation}
Both estimates are many orders of magnitude larger than the typical spin-down ages of radio pulsars. We would, therefore, expect 
Ohmic decay and Hall term to be negligible for the evolution of the plasma's magnetic field. In this idealised case, which is 
commonly used to approximate astrophysical or laboratory plasmas, the induction equation reduces to
\begin{align}
	 \partial_t B^i &= \epsilon^{ijk} \nabla_j \epsilon_{klm} \left( v_\rp^l B^m \right).
		\label{eqn-NormalMHDInductionIdeal}
\end{align}
Using Equation \eqref{eqn-MaxwellB}, we can rewrite the last expression and simplify the result using the standard Lie derivative,
\begin{align}
	 \partial_t B^i + \mathcal{L}_{v_\rp} B^i = - B^i \nabla_j v_\rp^j.
		\label{eqn-Lie2}
\end{align}
The left-hand side describes how the magnetic field vector, $B^i$, is transported with the fluid flow, $v_\rp^i$. Taking into account 
that the mass of the proton plasma is conserved, we use Equation \eqref{eqn-Continuity} to further simplify,
\begin{align}
	 \frac{\partial}{\partial t} \left( \frac{B^i}{\rho_\rp} \right) 
		+ \mathcal{L}_{v_\rp} \left( \frac{B^i}{\rho_\rp} \right) = 0.
		\label{eqn-Lie4}
\end{align}
This implies that the magnetic field is moving with the fluid, i.e. the fluxlines are frozen into the proton plasma. 

As soon as Ohmic and Hall terms play a role for the dynamics, this frozen-in condition is destroyed and field 
lines are no longer forced to follow the protons. In particular, if Ohmic decay characterised by the conductivity, $\sigma_\re$, 
is included, the induction equation resembles a diffusion equation. It encodes how the magnetic field lines diffuse through the 
fluid and reconnect, leading to the decay of magnetic energy as discussed in Section \ref{subsec-MHDEnergy}. If the 
Hall term is present but Ohmic decay is negligible, the induction equation reduces to
\begin{align}
	 \partial_t B^i &= \epsilon^{ijk} \nabla_j \epsilon_{klm} \left( v_\re^l B^m \right).
		\label{eqn-NormalMHDInductionHall}
\end{align}
In contrast to Equation \eqref{eqn-NormalMHDInductionIdeal}, the electron velocity enters the magnetic evolution law. This 
implies that the relative motion between electrons and protons becomes important and the magnetic field is frozen into the 
electron fluid. The Hall term in Equation \eqref{eqn-NormalMHDInduction} itself is not dissipative but may act to redistribute
magnetic energy from large scales to smaller ones, where it can decay ohmically. Many studies of the induction equation's non-linear 
behaviour are based on results from hydrodynamic turbulence \citep{Kolmogorov1941}, as it has several similarities with the 
vorticity equation of a viscous fluid \citep{Goldreich1992}. However, recent numerical simulations in the context of neutron 
stars have shown no evidence of strong cascading behaviour. Instead the Hall cascade appears to be saturated at long 
lengthscales \citep{Pons2010, Gourgouliatos2014, Wood2015}.

\subsection{Magnetic energy in standard MHD}
\label{subsec-MHDEnergy}

The conservative and dissipative nature of the different pieces in Equation \eqref{eqn-NormalMHDInduction} is illustrated 
by considering the evolution of the magnetic energy. In order to compare the standard MHD plasma with the superconducting 
mixture later on, we calculate the magnetic energy associated with the work done by the Lorentz force. In its standard form,
the Lorentz force density, given by
\begin{equation}
	F_\rL^i = \frac{1}{4 \pi} \left[ B_j \nabla^j B^i - \frac{1}{2} \nabla^i \left( B_k B^k \right) \right],
		\label{eqn-LorentzForce}
\end{equation}
is composed of a tension and a pressure term. The work is obtained by calculating the product with the position vector, 
$r^i$, and integrating over the volume, $V$. Using the product rule and Equation \eqref{eqn-MaxwellB}, we arrive at
\begin{align}
	W_\rL &= \int r_i F_\rL^i \, \rd V = \frac{1}{4 \pi} \int \nabla^i \left( r_j B^j B_i - 
		 \frac{1}{2} B^2 r_i \right) \rd V \nonumber \\[1.2ex]
	&- \frac{1}{4 \pi} \int \left( B^i B_j \nabla^j r_i - \frac{1}{2} B^2 \nabla^i r_i \right) \rd V.
		\label{eqn-Energy}
\end{align}
The total gradient term can be rewritten as a surface integral using Gauss' theorem. As no discontinuities are present at
the boundary of the plasma region, we can push the integration radius to infinity. Provided that the magnetic induction
vanishes at infinity, the surface contribution is zero. The integrand of the second term in Equation \eqref{eqn-Energy} 
simplifies to the well-known magnetic energy density
\begin{align}
	W_\rL &= \int \frac{B^2} {8 \pi} \, \rd V =  \int \cE_{\rm mag} \, \rd V.
\end{align}
Changes in the magnetic energy are, thus, determined by
\begin{equation}
	 \frac{\partial \cE_{\rm mag}}{\partial t} = \frac{\partial}{\partial t} \left( \frac{B^2}{8\pi} \right) 
		= \frac{B_i}{4\pi}  \frac{\partial B^i}{\partial t}. 
		\label{eqn-Energychange}
\end{equation}
Calculating the product of the induction equation with $B_i$ and using the product rule to rewrite the result, we find
\begin{align}
	\frac{\partial \cE_{\rm mag}}{\partial t}  &= \frac{1}{4 \pi} \, \epsilon^{isp} (\nabla_s B_p) 
		\bigg[ \epsilon_{ijk} v_\rp^j B^k - \frac{c^2}{4 \pi \sigma_\re} \,
		\epsilon_{ijk} (\nabla^j B^k) \nonumber \\[1.2ex]
		&- \frac{m c}{4 \pi e \rho_\rp} \, \epsilon_{ijk} \epsilon^{jlm} ( \nabla_l B_m ) B^k \bigg] 
		- \nabla^i \Sigma_i.
		\label{eqn-MagneticEnergy}
\end{align}
The last term contains all contributions that can be written as a divergence. After integrating over the volume, it is 
possible to convert this part into a surface integral using Gauss' theorem. Additionally, the third term has to be zero
due to the properties of the Levi-Civita tensor. As expected, the Hall term is conservative and does not contribute to 
the change in the magnetic energy density. Using Amp\`ere's law \eqref{eqn-AmpereLawStandard} and the generalised Ohm's 
law \eqref{eqn-OhmsLaw}, the remaining terms are simplified to
\begin{align}
	\frac{\partial \cE_{\rm mag}}{\partial t} &=  \frac{1}{c} \, J^i \epsilon_{ijk} \, v^j_\rp B^k - \frac{J^2}{\sigma_\re}
		\nonumber \\[1.4ex]
		&- \nabla^i \left[\frac{c}{4 \pi} S_i - \frac{m_\re}{e} \left( \tilde{\mu}_\re + \Phi \right)  J_i \right],
		\label{eqn-MagneticEnergy2}
\end{align}
where $S^i = \epsilon^{ijk} E_j B_k$ is the Poynting vector. Equation \eqref{eqn-MagneticEnergy2} clearly shows that any 
resistive plasma is subject to the decay of magnetic energy due to Ohmic diffusion and the energy loss is proportional to $J^2$. 
The inertial term vanishes if the protons are not able to move, which is, for example, the case in a standard metal, or the 
macroscopic current and the proton velocity are aligned.

%%%%%%%%%%%%%%%%%%%%%%%%%%%%%%%%%%%%%%%%%%%%%%%%%%%%%%%%%%%%%%%%%%%%%%%%%%%%%%%%%%
%%%%%%%%%%%%%%%%%%%%%%%%%%%%%%%%%%%%%%%%%%%%%%%%%%%%%%%%%%%%%%%%%%%%%%%%%%%%%%%%%%

\section{Magnetic field evolution in superconducting neutron stars}
\label{sec-SuperconMHD}

\subsection{The coupling force: `standard' resistivity}
\label{subsec-CouplingForce}

We now return to the question of magnetic field evolution in the superconducting outer neutron star core. In order to apply a 
formalism similar to the resistive MHD discussion, we need to determine the forces, $F^i_\re$, exerted on the electron component 
by the various fluid constituents and the vortices/fluxtubes. However, this is where things get complicated. Due to the multi-fluid
nature of the superfluid/superconducting mixture, there are not simply two components coupled by a single resistive force. We
could imagine a variety of ways for the components to interact with each other ranging from electron scattering 
\citep{Sauls1982, Alpar1984, Andersson2006b} and vortex-fluxtube interactions \citep{Ruderman1998, Jahan-Miri2000, Link2003} to 
shear or bulk viscosity \citep{Andersson2005, Shternin2008, Manuel2013}. Choosing a more pedagogical approach to our problem, we 
pick one specific mechanism, determine how it affects the electrons on mesoscopic scales and translate this into a macroscopic 
picture. While we won't provide a complete picture of the magnetic field evolution in the core, this method provides more insight 
to how different mechanisms could play a role. 

We keep in mind that the magnetic field is locked to the superconducting fluxtubes and their motion determines the evolution of 
the magnetic field. We, therefore, consider the scattering of electrons off the vortex/fluxtube magnetic fields as a source of
mutual friction. This `standard' resistive coupling in a superfluid/superconducting mixture, first discussed by \citet{Alpar1984}, 
results in two forces acting on the electrons,
\begin{equation}
	 F^i_{\re} = F^i_{\rm pe} + F^i_{\rm ne} \approx F^i_{\rm pe}.
\end{equation}
We neglect the contribution from electrons scattering off the neutron vortices because $\cN_\rp \gg \cN_\rn$. This implies that 
electron-fluxtube interactions are markedly more common and, thus, dominate the electron coupling. The macroscopic force, $F^i_{\re}$, is 
obtained by multiplying the electron drag force, $f^i_{\rm d}$, exerted on a single fluxtube, with the fluxtube density, $\cN_\rp$,
\begin{equation}
	F^i_{\rm pe} = \cN_\rp  f^i_{\rm d}  = \cN_\rp \rho_\rp \kappa \cR \left(v^i_\re -u^i_\rp\right).
		  \label{eqn-DragElectron}
\end{equation}
The dimensionless drag coefficient, $\cR$, contains all the information about the coupling on mesoscopic scales and $u^i_\rp$ is 
the velocity of a single fluxtube. 

In order to determine an evolution equation for the macroscopic magnetic field in superconducting 
matter, we have to eliminate any quantities from Equation \eqref{eqn-DragElectron} that are defined on mesoscopic lengthscales. 
Hence, the next step is to rewrite the fluxtube velocity, $u^i_\rp$, in terms of the macroscopic fluid variables. This can be 
achieved by using the force balance for an individual fluxtube, an approach introduced by \citet{Hall1956} for the description of 
superfluid helium. We have
\begin{equation}
	\sum f^i = f^i_{\rm d} + f^i_{\rm M} + f^i_{\rm t} + f^i_{\rm em} =0,
		\label{eqn-ForceBalanceFluxtube}
\end{equation}
where the fluxline inertia is neglected. Our force balance equation includes the electron drag force given above, the Magnus force, 
$f^i_{\rm M}$, the tension force, $f^i_{\rm t}$, and the electromagnetic Lorentz force, $f^i_{\rm em}$. The different forces have been 
calculated by \citet{Glampedakis2011a} and are given by
\begin{align}
	f^i_{\rm M} &= - \rho_\rp \kappa \, \epsilon^{ijk} \hat{\kappa}_j^\rp \left(v_k^\rp -u_k^\rp \right), 
		\label{eqn-MagnusForce} \\[1.8ex]
	f^i_{\rm t} &= \frac{H_{\rc 1} \kappa}{4 \pi a_\rp} \, \hat{\kappa}^j_\rp \nabla_j \hat{\kappa}^i_\rp, 
		\label{eqn-TensionForce}
\end{align}
where $H_{\rc 1}$ is the lower critical field for superconductivity \citep{Tilley1990}, and 
\begin{equation}
	f^i_{\rm em} = \rho_\rp \kappa \, \epsilon^{ijk} \hat{\kappa}_j^\rp w_k^{\rp \re}.
		\label{eqn-EmForce}
\end{equation}
Note at this point that we are interested in the linear analysis of one specific resistive mechanism. For this reason, our force 
balance \eqref{eqn-ForceBalanceFluxtube} does not include a `pinning' force, resulting from the magnetic short-range interaction 
between the two arrays \citep{Ruderman1998, Jahan-Miri2000, Link2003, Glampedakis2011}. 

Calculating repeated cross products of the force balance equation with $\hat{\kappa}^i_\rp$, pointing 
along the local orientation of a fluxtube, it is possible to express the mesoscopic fluxtube velocity in terms of the averaged
fluid velocities,
\begin{equation}
	 u _\rp^i = v_\re^i + \frac{1}{1+\cR^2} \left( \cR f_\star^i + \epsilon^{ijk} \hat{\kappa}_j^\rp f^{\star}_k \right),
		\label{eqn-VelocityProtons1}
\end{equation}
where
\begin{equation}
	f_{\star}^i = \epsilon^{ijk} \hat{\kappa}^\rp_j w^{\rm ep}_k  + \frac{1}{\rho_\rp \kappa} 
		\left(f^i_{\rm t} + f^i_{\rm em} \right).
		\label{eqn-ForceStar1}
\end{equation}
Combining the previous relations, we observe that the first term in Equation \eqref{eqn-ForceStar1} and the electromagnetic force, 
$f^i_{\rm em}$, cancel each other. Then, the effective force, $f_{\star}^i$, is equivalent to the fluxtube tension, 
\begin{equation}
	f_{\star}^i = \frac{1}{\rho_\rp \kappa} \, f^i_{\rm t} 
		= \frac{H_{\rc 1}}{4 \pi a_\rp \rho_\rp} \hat{\kappa}^j_\rp \nabla_j \hat{\kappa}^i_\rp.
		\label{eqn-ForceStar2}
\end{equation}
Substituting Equations \eqref{eqn-VelocityProtons1} and \eqref{eqn-ForceStar2} back into Equation \eqref{eqn-DragElectron} 
finally gives for the macroscopic drag force,
\begin{equation}
	F_{\re}^i = - \frac{H_{\rm c1} \phi_0 \cN_\rp \cR}{4\pi (1+\cR^2)} \left(\cR \, \hat{\kappa}_\rp^j 
		\nabla_j \hat{\kappa}^i_\rp + \epsilon^{ijk} \hat{\kappa}_j^\rp \hat{\kappa}^l_\rp \nabla_l \hat{\kappa}_k^\rp
		\right).
		\label{eqn-ForceElectronsSC}
\end{equation}
For a straight fluxtube array, the tension force and, thus, the electron coupling would vanish.

\subsection{The superconducting induction equation}
\label{subsec-SuperconInduction}

Having determined the force, $F^i_\re$, exerted on the electron component due to scattering off the fluxtube magnetic fields,
we use Equation \eqref{eqn-ElectricField} for the macroscopic electric field to derive a generalised Ohm's law that is valid 
in the superfluid/superconducting mixture,
\begin{align}
	E^i &= -\frac{1}{c} \epsilon^{ijk} v_j^\re  B_k -\frac{m_\re}{e} \nabla^i \left ( \tilde{\mu}_\re + \Phi \right )
		+ \frac{H_{\rm c1} \phi_0 \cN_\rp}{4\pi c a_\rp \rho_\rp} \nonumber \\[1.2ex]
		&\times \frac{\cR}{1+\cR^2}  \left( \cR \, \hat{\kappa}_\rp^j 
		\nabla_j \hat{\kappa}^i_\rp + \epsilon^{ijk} \hat{\kappa}_j^\rp \hat{\kappa}^l_\rp \nabla_l \hat{\kappa}_k^\rp
		\right). 
		\label{eqn-ElectricCurrentSC2}
\end{align}
With Faraday's law \eqref{eqn-Faraday}, we obtain a superconducting induction equation describing the evolution of the
macroscopic magnetic field in the outer neutron star core,
\begin{align}
	\partial_t B^i &= \epsilon^{ijk} \nabla_j \bigg[ \epsilon_{klm} \left( v^l_\re B^m \right)- \frac{H_{\rm c1} \phi_0 \cN_\rp}
		{4\pi a_\rp \rho_\rp}  \nonumber \\[1.2ex] &\times \frac{\cR}{1+\cR^2} \left( \cR \, \hat{\kappa}_\rp^l \nabla_l 
		\hat{\kappa}^\rp_k + \epsilon_{klm} \hat{\kappa}^l_\rp \hat{\kappa}^s_\rp \nabla_s  \hat{\kappa}_\rp^m
		\right) \bigg].
		\label{eqn-InductionFull}
\end{align}
Let us specify the lower critical field, $H_{\rm c1}$, to simplify this expression further \citep{Tilley1990}.
The field is related to the energy per unit length of the fluxtube, $\mathcal{E}_\rp$, via
\begin{equation}
	H_{\rm c1} = \frac{4 \pi \mathcal{E}_\rp}{\phi_0}.
		\label{eqn-LowerCriticalField}
\end{equation}
The fluxtube energy, on the other hand, is determined by the characteristic lengthscales of the superconducting phase and
given by
\begin{equation}
	\mathcal{E}_\rp = \left( \frac{\phi_0} {4 \pi \lambda_\star} \right)^2 \text{ln} \left( \frac{\lambda_*}{\xi_\rp}\right).
		\label{eqn-SelfEnergy1}
\end{equation}
Here, $\xi_\rp$ denotes the proton coherence length and the effective London penetration depth was defined in Equation 
\eqref{eqn-LondonDepth}. According to \citet{Tinkham2004}, one can estimate $\text{ln} \left( \lambda_*/ \xi_\rp\right) 
\approx 2$, which leads to
\begin{equation}
	\mathcal{E}_\rp \approx \frac{\rho_\rp \kappa^2} {2 \pi} \frac{1-\en}{1 -\en-\ep}
		\approx \frac{\rho_\rp \kappa^2} {2 \pi} \frac{m}{m_\rp^*} .
		\label{eqn-SelfEnergy2}
\end{equation}
Combining the previous equations, we find
\begin{align}
	\partial_t B^i &= \epsilon^{ijk} \nabla_j \bigg[ \epsilon_{klm} \left( v^l_\re B^m \right) - \frac{\kappa  \phi_0 \cN_\rp}
		{2 \pi} \frac{m}{m_\rp^*}  \nonumber \\[1.2ex] &\times \frac{\cR}{1+\cR^2} \left( \cR \, \hat{\kappa}_\rp^l \nabla_l 
		\hat{\kappa}^\rp_k + \epsilon_{klm} \hat{\kappa}^l_\rp \hat{\kappa}^s_\rp \nabla_s  \hat{\kappa}_\rp^m
		\right) \bigg],
		\label{eqn-InductionFullFinal}
\end{align}
which is the main result of our paper.

\subsection{A simplified set of equations and the field evolution timescales}
\label{subsec-SimplifiedSet}

At this point, it seems natural to make several assumptions about the actual physics of the multi-fluid mixture inside a 
neutron star in order to find a simplified version of Equation \eqref{eqn-InductionFullFinal}. As discussed previously, the main 
contribution to the macroscopic magnetic induction is given by the fluxtubes. In this case, we can neglect the weak London
field and the superconducting Amp\`ere law \eqref{eqn-AmpereLaw} dictates that the protons and electrons are co-moving on large 
scales, i.e. $v^i_\rp \approx v^i_\re$, and the macroscopic current vanishes. This also implies that the local direction of the 
fluxtube array is aligned with the direction of the magnetic field because
\begin{equation}
		B^i = B \hat{B}^i \approx \cN_\rp \phi_0 \hat{\kappa}^i_\rp \qquad \text{gives} \qquad \hat{B}^i \approx \hat{\kappa}^i_\rp.
\end{equation}
Using these simplifications to rewrite the force on the electron fluid, we obtain
\begin{equation}
	F_{\re}^i \approx - \frac{H_{\rm c1} B \cR}{4\pi (1+\cR^2)} \left(\cR \hat{B}^j 
		\nabla_j \hat{B}^i + \epsilon^{ijk} \hat{B}_j \hat{B}^l \nabla_l \hat{B}_k
		\right).
		\label{eqn-ForceElectronsSCSimplifieda}
\end{equation}
%Show that there is a conservative and a dissipative contribution ...
The induction equation, on the other hand, reduces to
\begin{align}
	\partial_t B^i &\approx \epsilon^{ijk} \nabla_j \bigg[ \epsilon_{klm} \left( v^l_\rp B^m \right) - \frac{\kappa B}
		{2 \pi} \frac{m}{m_\rp^*} \nonumber \\[1.2ex] &\times \frac{\cR}{1+\cR^2} \left( \cR \hat{B}^l \nabla_l 
		\hat{B}_k + \epsilon_{klm} \hat{B}^l \hat{B}^s \nabla_s  \hat{B}^m
		\right) \bigg].
		\label{eqn-InductionSimplifieda}
\end{align}
We will compare this form of the superconducting induction equation with the standard MHD result in Equation 
\eqref{eqn-NormalMHDInduction}. As in the resistive MHD case, we can extract two timescales,
\begin{equation}
	 \tau_{\rm diss} = \frac{2 \pi L^2 (1+\cR^2)} {\kappa \cR}\, \frac{m_\rp^*} {m}
		\label{eqn-tau1eqn}
\end{equation}
and 
\begin{equation}
	 \tau_{\rm cons} =  \frac{\tau_{1}}{\cR} = \frac{2 \pi L^2(1+\cR^2)} {\kappa \cR^2} \, \frac{m_\rp^*} {m},
		\label{eqn-tau2eqn}
\end{equation}
where $L$ is again the characteristic lengthscale over which the magnetic field changes. The naming convention of the 
two timescales might seem arbitrary at this point, but our choice will become clear later on.

We can estimate these two timescales provided the strength of the mutual friction is known. A method to calculate the 
dimensionless drag parameter, $\cR$, for the coupling of relativistic electrons and a single fluxtubes is discussed in 
\citet{Alpar1984} (see also \citeauthor{Sauls1982} \citeyear{Sauls1982}; \citeauthor{Andersson2006b} 
\citeyear{Andersson2006b})\footnote{Note that according to \citet{Jones2006}, magnetic scattering off individual 
fluxtubes is suppressed for very large fluxtube densities. Instead electron scattering by a cluster of fluxtubes dominates
the coupling, leading to a much smaller drag coefficient, $\cR$.}.
The authors give the relaxation timescale, $\tau_{\rm pe}$, for the `resistive' interaction and include effects caused 
by the finite fluxtube size and the increase in moment of inertia due to the coupling of electrons and protons on much 
shorter timescales. The relaxation timescale is related to the drag coefficient via
\begin{equation}
	\cR = \left(\kappa \, \cN_\rp \tau_{\rm pe} \right)^{-1},
\end{equation}
which leads to the following numerical estimate
\begin{equation} \hspace{-0.01cm}
	\cR \approx 1.9 \times 10^{-4} \left( \frac{m}{m_\rp^*} \right)^{1/2} \rho_{14}^{1/6} \left(\frac{x_\rp}{0.05} \right)^{1/6}.
\end{equation}
This gives $\cR \ll 1$ and implies that the standard friction mechanism is rather weak. Adopting this limit, we can 
approximate for the neutron star core,
\begin{equation}
	 \tau_{\rm diss} \approx 3.1 \times 10^{11} L_6^2 \, \rho_{14}^{-1/6} \left( \frac{x_\rp}{0.05}\right)^{-1/6} \, \yr
		\label{eqn-tau1estimate}
\end{equation}
and 
\begin{equation}
	 \tau_{\rm cons} \approx 1.3 \times 10^{15} L_6^2 \, \rho_{14}^{-2/6} \left(\frac{x_\rp}{0.05}\right)^{-2/6} \, \yr.
		\label{eqn-tau2estimate}
\end{equation}
We have used Equation \eqref{eqn-envalue} to estimate the effective proton mass, i.e. the entrainment parameter. 

\subsection{Flux-freezing and magnetic energy}
\label{subsec-SuperconEnergy}

For conventional electron-fluxtube coupling, the timescales for the magnetic field evolution are rather long and the
dynamics of the macroscopic induction are dominated by the inertial term in the induction equation. In the weak mutual 
friction limit, we are, thus, left with an equation that is equivalent to the ones discussed in Subsection 
\ref{subsec-MHDFreezing}. The magnetic field in the superconducting sample is frozen to proton fluid, which implies that 
the superconducting fluxtubes are locked to the proton plasma, i.e. $v_\rp^i \approx u_\rp^i$. Hence, electrons, protons 
and fluxtubes are comoving on large scales, which is different to the weakly resistive case of standard MHD, where the 
relative motion between the charged particles was important.

In order to determine whether the additional terms in Equation \eqref{eqn-InductionSimplifieda} are conservative or 
dissipative and which timescale dominates, we discuss the evolution of the superconducting magnetic energy. As before, 
we evaluate the energy associated with the work done by the magnetic force. However, in a superfluid/superconducting mixture 
the standard Lorentz force \eqref{eqn-LorentzForce} has to be changed accordingly. For a non-rotating star in the absence 
of entrainment the total magnetic force is given by \citep{Easson1977, Akgun2008, Glampedakis2011a, Lander2013} 
\begin{equation}
	F_{\rm mag}^i = \frac{1}{4 \pi} \left[ B_j \nabla^j H_{\rm c1}^i -  
		\nabla^i \left( \rho_\rp B \frac{\partial H_{\rm c1}}{\partial \rho_\rp} \right) \right]
		\label{eqn-SuperconForce},
\end{equation}
where $H_{\rm c1}^i = H_{\rm c1} \hat{B}^i$. This expression also contains a tension and a pressure term but both scale with 
$H_{\rm c1}^i $ and $B^i$ instead of $B^2$. The work associated with this force is given by 
\begin{align}
	W_{\rm mag} &=  \frac{1}{4 \pi} \int \nabla^i \left( r_j H_{\rm c1}^j B_i - 
		 \rho_\rp B \frac{\partial H_{\rm c1}}{\partial \rho_\rp} \, r_i \right) \rd V \nonumber \\[1.2ex]
	&- \frac{1}{4 \pi} \int \left( H_{\rm c1}^i B_j \nabla^j r_i -  \rho_\rp B \frac{\partial H_{\rm c1}}{\partial \rho_\rp} 
		\nabla^i r_i \right) \rd V,
		\label{eqn-EnergySuperconA}
\end{align}
where we have used the product rule and Equation \eqref{eqn-MaxwellB}. Similar to the standard MHD case, the first term can be 
rewritten using Gauss' theorem. However, in the superconducting outer core, this contribution does not simply vanish because 
discontinuities are likely to be present at the fluids' boundaries. The dynamics that might arise due the presence of a current 
sheet at the crust-core interface or the type-II to type-I transition region in the neutrons star's inner core are only poorly 
understood and significantly complicate the problem. Incorporating these interfaces would require a much more detailed understanding 
of the microphysics involved. In the following, we, therefore, omit a discussion of the surface terms and focus on the much simpler 
problem of magnetic field evolution in the bulk fluid. 

Taking into account that $H_{\rm c1}$ is a function linear in $\rho_\rp$, the second integral in Equation \eqref{eqn-EnergySuperconA} 
reduces to \begin{align}
	W_{\rm mag, bulk} & = \int \frac{H_{\rm c1} B} {2 \pi} \, \rd V = \int \cE_{\rm mag,sc} \, \rd V.
\end{align}
Taking the time derivative of the magnetic energy density gives two contributions
\begin{equation}
	 \frac{\partial  \cE_{\rm mag,sc}}{\partial t} = \frac{B}{2 \pi} \frac{\partial H_{\rm c1}} {\partial t}  
		+  \frac{H_{\rm c1}}{2 \pi} \hat{B}_i \frac{\partial B^i}{\partial t}. 
		\label{eqn-EnergychangeSC}
\end{equation}
Comparison with the corresponding expression of standard MHD given in Equation \eqref{eqn-Energychange} shows that the superconducting 
nature of the mixture gives rise to a new contribution for the change in energy density. In contrast to resistive MHD, the evolution of 
matter and the magnetic induction are no longer decoupled in the condensate. Equation \eqref{eqn-EnergychangeSC} demonstrates that 
modifying the properties of the superconductor, such as the lower critical field, $H_{\rm c1}$, alters the magnetic energy. This implies 
that an evolving matter configuration can be closely linked to a changing magnetic field.

The second term in Equation \eqref{eqn-EnergychangeSC} is similar to the result for normal conducting matter. Calculating the product of 
the induction equation \eqref{eqn-InductionSimplifieda} with $\hat{B}_i$ and using the product rule to simplify, we find 
\begin{align}
	\hat{B}_i \frac{\partial B^i}{\partial t} &=  \epsilon^{isp} (\nabla_s \hat{B}_p) \bigg[ \epsilon_{ijk} v_\rp^j B^k - 
		\frac{\kappa B \cR^2} {2 \pi (1+\cR^2)} \frac{m}{m_\rp^*} \, \hat{B}^l \nabla_l  \hat{B}_i  \nonumber \\[1.2ex]
		&-  \frac{\kappa B \cR} {2 \pi (1+\cR^2)} \frac{m}{m_\rp^*} \, \epsilon_{ijk}  \hat{B}^j  \hat{B}^l \nabla_l  \hat{B}^k \bigg] 
		 - \nabla^i \Sigma_i.
		\label{eqn-SuperconMagneticEnergy}
\end{align}
As before, $\Sigma^i$ denotes the divergence terms. This equation bears some resemblance with the result \eqref{eqn-MagneticEnergy} 
found in standard MHD and we would equivalently expect to obtain a conservative and a dissipative contribution. In order to determine which 
of the terms are nonzero or zero, we rewrite the tension using the following identity
\begin{align}
		\hat{B}^l \nabla_l  \hat{B}_i = \epsilon_{ijk}  \epsilon^{jlm} ( \nabla_l \hat{B}_m ) \hat{B}^k.
			\label{eqn-TensionIdentity}
\end{align}
The second term in Equation \eqref{eqn-SuperconMagneticEnergy} is, thus, proportional to
\begin{align} 
	\epsilon^{isp} (\nabla_s \hat{B}_p)  \hat{B}^l \nabla_l  \hat{B}_i = 
		\epsilon^{isp} (\nabla_s \hat{B}_p)  \, \epsilon_{ijk}  \epsilon^{jlm} (\nabla_l \hat{B}_m) \hat{B}^k.
		\label{eqn-SuperconHall}
\end{align}
Analogous to the Hall term of standard resistive MHD, this vanishes due to the properties of the antisymmetric Levi-Civita tensor. 
Thus, the second term in the superconducting induction equation \eqref{eqn-InductionSimplifieda} is conservative and does not 
modify the total magnetic energy of the superconducting mixture. The last term in Equation \eqref{eqn-SuperconMagneticEnergy}, on 
the other hand, is proportional to
\begin{align}
	 \epsilon^{isp} (\nabla_s \hat{B}_p) \, \epsilon_{ijk}  \hat{B}^j  \hat{B}^l \nabla_l  \hat{B}^k 
	 	= \mathcal{J}^i \epsilon_{ijk}  \hat{B}^j  \epsilon^{klm}  \mathcal{J}_l \hat{B}_m,
\end{align}
where we defined the vector
\begin{equation}
	\mathcal{J}^i \equiv \epsilon^{ijk} \nabla_j \hat{B}_k.
\end{equation}
Rewriting the remaining two Levi-Civita tensors in terms of Kronecker deltas, $\delta^i_j$, gives the projection
\begin{align}
	\mathcal{J}^i  \, \epsilon_{ijk}  \hat{B}^j  \epsilon^{klm}  \mathcal{J}_l \hat{B}_m
		= \mathcal{J}_i  \mathcal{J}^i  - \left( \mathcal{J}^i \hat{B}_i \right) \left( \mathcal{J}^j \hat{B}_j\right).
		\label{eqn-IdentityOhm}
\end{align}
Decomposing the vector $\mathcal{J}^i$ into a component parallel to $\hat{B}_i$ and one perpendicular to the magnetic field 
direction, i.e. $\mathcal{J}^i = \mathcal{J}_{\parallel} \hat{B}_i + \mathcal{J}_{\perp}^i$, we see that Equation 
\eqref{eqn-IdentityOhm} only depends on the component of $\mathcal{J}^i$ that is perpendicular to $\hat{B}_i$,
\begin{align}
	\mathcal{J}_i  \mathcal{J}^i  - \left( \mathcal{J}^i \hat{B}_i \right) \left( \mathcal{J}^j \hat{B}_j\right)
	= \mathcal{J}_{\perp}^2.
		\label{eqn-IdentityOhm2}
\end{align}
Similar to the Ohmic term in standard MHD, we also retain a dissipative contribution to the total magnetic energy in the case of 
superconducting MHD. It is given by
\begin{align}
	\frac{\partial  \cE_{\rm mag,sc}}{\partial t}	&=   \frac{B}{2 \pi} \frac{\partial H_{\rm c1}} {\partial t}  
		+ \frac{H_{\rm c1}}{2 \pi} \mathcal{J}^i_\perp \epsilon_{ijk} v_\rp^j B^k \nonumber \\[1.4ex]
		&-  \frac{H_{\rm c1} \kappa B \cR} {4 \pi^2 (1+\cR^2)} \frac{m}{m_\rp^*} \,\mathcal{J}_{\perp}^2
		- \frac{H_{\rm c1}}{2 \pi} \nabla^i \Sigma_i.
		\label{eqn-SuperconMagneticEnergyFinal}
\end{align}

Having calculated the change in the magnetic energy density, we can associate the timescale $\tau_{\rm diss}$, given in Equation 
\eqref{eqn-tau1estimate}, with a dissipative mechanism. Comparing the numerical estimate to the Ohmic decay timescale 
\eqref{eqn-Ohmicestimate}, we observe that the resistive coupling in a superfluid/superconducting mixture acts on a timescale,
which is two orders of magnitude smaller than the standard MHD diffusion,
\begin{equation}
	  \frac{\tau_{\rm diss}}{\tau_{\rm Ohm}} \approx 1.3 \times 10^{-2} \, 
		T_8^2  \, \rho_{14}^{-5/3} \left( \frac{x_\rp}{0.05}\right)^{-5/3}.
\end{equation}
On the other hand, the timescale \eqref{eqn-Hallestimate} for the Hall evolution in a normal conducting plasma can be compared 
to $\tau_{\rm cons}$ in Equation \eqref{eqn-tau2estimate}. In contrast to the standard MHD case, the conservative timescale emerging 
from the superconducting induction equation is several orders of magnitude larger than the Hall timescale, 
\begin{equation}
	  \frac{\tau_{\rm cons}}{\tau_{\rm Hall}} \approx 6.8 \times 10^{4} \, 
		B_{12} \, \rho_{14}^{-4/3} \left( \frac{x_\rp}{0.05}\right)^{-4/3}.
\end{equation}
We also note that due to the dependence on the dimensionless drag coefficient, $\cR$, the dissipative term in the superconducting 
induction equation governs the magnetic field evolution, whereas in standard MHD the conservative Hall term acts on shorter 
timescales. In the latter case, the order of the two timescales is necessary for any cascading behaviour to take place; the Hall 
term drives the magnetic field to shorter lengthscales, where it can decay ohmically. However, if diffusion is dominating the 
evolution, the redistribution of magnetic energy will happen on much longer timescale not causing a cascade. Hence, despite the 
close similarities between the conservative terms in Equation \eqref{eqn-InductionSimplifieda} and standard MHD, we conjecture
that the analysis of the Hall cascade by \citet{Goldreich1992} is not transferable to the superconducting case.

%%%%%%%%%%%%%%%%%%%%%%%%%%%%%%%%%%%%%%%%%%%%%%%%%%%%%%%%%%%%%%%%%%%%%%%%%%%%%%%%%%
%%%%%%%%%%%%%%%%%%%%%%%%%%%%%%%%%%%%%%%%%%%%%%%%%%%%%%%%%%%%%%%%%%%%%%%%%%%%%%%%%%

\section{Discussion}
\label{sec-Discussion}

Strong magnetic fields are a key ingredient for many phenomena observed in neutron stars. Understanding the fields' long-term 
evolution might give insight into the `metamorphosis' between the different neutron star classes, the changing fields of 
standard radio pulsars or the high activity of magnetars. As the mechanisms causing field changes are only poorly understood, 
we revisited the question of magnetic field evolution and, in particular, discussed the influence of a superconducting component. 
Our aim was to rethink key notions of magnetohydrodynamics and develop a better intuition for the magnetic field evolution in 
a superconductor.

In this work, we used a multi-fluid formalism to describe the mixture in the outer neutron star core. The model introduced 
in Section \ref{sec-Background} translates the presence of mesoscopic vortices/fluxtubes into the large scale dynamics of 
the fluid. As an application of this framework, we analysed the conventional dissipative mechanism, i.e. the scattering of 
electrons off the fluxtube magnetic field. Based on the approach of standard resistive MHD, we combined a generalised Ohm's 
law with Faraday's law and the respective force on the electron fluid to derive a superconducting induction equation. 
Considering the London field as a negligible contribution led to a simplified equation, which should be applicable to most 
astrophysical scenarios. Caution is in order when discussing highly magnetised objects. For field strengths above the
upper critical field, $B > H_{\rm c2} \sim 10^{16} \, \G$ \citep{Tilley1990}, the superconducting state breaks down and our 
averaged formalism no longer applies. According to \citet{Goldreich1992}, ambipolar diffusion could potentially
become important in this regime and drive field decay on the order of typical magnetar ages.

To compare our new results for magnetic field evolution with the standard MHD case, the magnetic energies associated with the 
total magnetic forces were calculated. In our analysis, we significantly simplified the problem by omitting a detailed discussion 
of the surface terms, a key problem which needs to be addressed in future studies. This implies that effects originating at 
the crust-core interface or the type-II to type-I transition in the inner core, which could potentially drive the magnetic field 
evolution, are not taken into account. Instead, we focused on the evolution of the averaged magnetic field in the bulk.
We found that in the limit of weak mutual friction, the inertial term dominates the field evolution. The fluxtubes move with 
the proton fluid and the flux is, as in the standard MHD case, frozen to the charged particles. We additionally showed that 
the new induction equation contains a dissipative and a conservative contribution, similar to the Ohmic and the Hall term in 
normal conducting matter. However, the evolution timescales extracted from the superconducting induction equation for weak 
mutual friction, $10^{11} \, \yr$ and $10^{15} \, \yr$ respectively, are notably longer than the typical spin-down ages of 
neutron stars. We conclude that the conventional mutual friction mechanism cannot serve as an explanation for the field changes
in pulsars or the activity of magnetars, which would require timescales of order $10^7\, \yr$ and $10^4 \, \yr$, respectively. 
Simply increasing the strength of the mutual friction cannot provide a solution to this problem either. Due to the $\cR$-dependence 
of $\tau_{\rm diss}$, the minimum dissipation timescale one could obtain with a frictional mechanism of the form \eqref{eqn-DragElectron} 
is
\begin{equation}
	\tau_{\min} = \frac{4\pi L^2}{\kappa} \, \frac{m_\rp^*}{m} \approx 1.4 \times 10^{8} L_6^2 \, \rho_{14}^{-1/6} 
	\left( \frac{x_\rp}{0.05}\right)^{-1/6} \, \yr
\end{equation}
for $\cR =1$. We note at this point that all numerical estimates crucially depend on the lengthscale $L$. It can be identified 
with the curvature radius of the magnetic field and we chose the neutron star radius, $R$, to normalise the previous results. 
Recent work on field equilibria in superconducting neutron star cores by \citet{Lander2013} suggests that the field configuration 
actually supports structures on a shorter lengthscale of $L \approx 10^5 \, \cm$. Adopting such an estimate would reduce the characteristic 
timescales by two orders of magnitude. In particular, the minimum dissipation timescale, $\tau_{\min}$, would be shortened to a million 
years, which is closer to the timescales of astrophysical interest.

Our results additionally suggest that the highly conducting neutron star core might affect the crustal field and slow down its 
evolution. Making a precise statement at this point is, however, not possible due to the poorly known physics at the crust-core 
boundary. This transition is crucial in understanding how changes of the core magnetic field are translated to the crust. The 
analysis presented in this paper does, therefore, not reconcile the discrepancy between short crustal decay timescales 
\citep{Pons2009} and the much longer core evolution. In order to significantly reduce the latter different dissipative mechanisms 
have to be invoked. The typical candidate for strong coupling is vortex-fluxtube `pinning' due to the short-range magnetic 
interaction between the two arrays. While we did not address pinning specifically, discussing the vortex-fluxtube interaction 
would be the natural continuation of this paper as our prescription can deal with any coupling mechanism. Based on a mesoscopic 
description of the pinning process, one would have to determine how the coupling affects the electron fluid and substitute the 
respective force, $F_\re^i$, into the generalised Ohm's law. Determining the superconducting induction equation that would result 
from the pinning interaction will be left for future work.

% - Meissner effect comment: we use macroscopic magneto hydrodynamics for a superconducting bulk,
% Meissner effect itself is not explicitly included in our formalism, we expect it to work on a
% timescale that is much larger than our hydro timescales we are interested in, in our naive picture
% we consider the motion of fluxlines out of a fluid element due to dissipative interactions, in 
% principle the Meissner expulsion of fluxtubes is an instability type problem where at a critical
% $\kappa$, the system wants to move into a lower energy state, i.e. from a type-II to a type-I state,
% exactly what we might expect to happen at higher densities in the stars centre, this would lead to
% fluxexpulsion on large scales, but it is not clear what actually happens on microscopic scales

% - drag coefficient is not constant in space but depends on the density of the star

% - we neglect pinning but there might be an opposite regime where the pinning is strong (such as the example of nulling precession by
% Link) and the spin is the quantity that is mainly influenced and not the magnetic field, which would be the exact opposite to our 
% calculation where we assume that the spin is unaffected but the field decays

%%%%%%%%%%%%%%%%%%%%%%%%%%%%%%%%%%%%%%%%%%%%%%%%%%%%%%%%%%%%%%%%%%%%%%%%%%%%%%%%%%

\section*{Acknowledgements}

VG is supported by Ev. Studienwerk Villigst and NA acknowledges support from STFC in the UK.
KG is supported by the Ram\'{o}n y Cajal Programme of the Spanish Ministerio de Ciencia e 
Innovaci\'{o}n and by the German Science Foundation (DFG) via SFB/TR7.

%%%%%%%%%%%%%%%%%%%%%%%%%%%%%%%%%%%%%%%%%%%%%%%%%%%%%%%%%%%%%%%%%%%%%%%%%%%%%%%%%%
%%%%%%%%%%%%%%%%%%%%%%%%%%%%%%%%%%%%%%%%%%%%%%%%%%%%%%%%%%%%%%%%%%%%%%%%%%%%%%%%%%

% \bibliographystyle{ieeetr}
% \bibliographystyle{ieeetr}

\bibliographystyle{mn2e}

% \bibliography{library}

\bibliography{../../Bibliography/library}

%%%%%%%%%%%%%%%%%%%%%%%%%%%%%%%%%%%%%%%%%%%%%%%%%%%%%%%%%%%%%%%%%%%%%%%%%%%%%%%%%%

\end{document}